\documentclass[showpacs,amsmath,amssymb,aps,showkeys,floatfix,prd,a4paper]{revtex4}

\usepackage{dcolumn}
\usepackage{bm}
\usepackage{epsfig}
\usepackage{amsfonts}
\usepackage{amssymb,amscd}

\def\lsim{\raise0.3ex\hbox{$<$\kern-0.75em\raise-1.1ex\hbox{$\sim$}}}

\def\gsim{\raise0.3ex\hbox{$>$\kern-0.75em\raise-1.1ex\hbox{$\sim$}}}

\newcommand{\be}{\begin{equation}}

\newcommand{\ee}{\end{equation}}

\def\beq{\begin{equation}}

\def\eeq{\end{equation}}

\def\beqa{\begin{eqnarray}}

\def\eeqa{\end{eqnarray}}

\newcommand{\ba}{\begin{eqnarray}}

\newcommand{\ea}{\end{eqnarray}}

\def\gappeq{\mathrel{\rlap {\raise.5ex\hbox{$>$}}

{\lower.5ex\hbox{$\sim$}}}}

\def\lappeq{\mathrel{\rlap{\raise.5ex\hbox{$<$}}

{\lower.5ex\hbox{$\sim$}}}}

\def\Toprel#1\over#2{\mathrel{\mathop{#2}\limits^{#1}}}

\begin{document}

\title{Track signals at IceCube from subleading channels}

\author{Reinaldo {\sc Francener}}
\email{reinaldofrancener@gmail.com}
\affiliation{Instituto de Física Gleb Wataghin - UNICAMP, 13083-859, Campinas, SP, Brazil. }

\author{Victor P. {\sc Gon\c{c}alves}}
\email{barros@ufpel.edu.br}
\affiliation{Institute of Physics and Mathematics, Federal University of Pelotas, \\
  Postal Code 354,  96010-900, Pelotas, RS, Brazil}

\author{Diego R. {\sc Gratieri}}
\email{drgratieri@id.uff.br}
\affiliation{Escola de Engenharia Industrial Metal\'urgica de Volta Redonda,
Universidade Federal Fluminense (UFF),\\
 CEP 27255-125, Volta Redonda, RJ, Brazil}
\affiliation{Instituto de Física Gleb Wataghin - UNICAMP, 13083-859, Campinas, SP, Brazil. }

\begin{abstract}
Tracks events at the IceCube Observatory are characterized by an energetic muon  crossing several kilometers before decaying. 
Such muons are dominantly produced in charged  current (CC) muon neutrino - hadron interactions. However, muons are also produced through the $W$ - boson production and in the decay of tau leptons and heavy mesons created in neutral and charged current interactions induced by all neutrino flavors. In this paper, we investigate the contribution of these subleading channels to events characterized as tracks at the IceCube. Our results indicate that these channels correspond to a non - negligible fraction of the HESE track events. 
In addition, we show that its contributions are concentrated in muons that are less energetic than those arising from muonic neutrino CC interactions for the same visible energies of the process. Finally, we investigate the impact of these additional channels on the description of the astrophysical neutrino flux, and we find that the inclusion of these subleading processes are important in determining the parameters of the astrophysical neutrino flux.

\end{abstract}



\maketitle

\vspace{1cm}

\section{Introduction}

One of the main goals of the IceCube observatory is the observation and description of ultrahigh-energy astrophysical neutrinos. Since the beginning of its operation, the associated neutrino flux have been constrained in different analyses, considering new data and distinct topologies \cite{HESE1,HESE2,tracksTotal,tracksNorte,cascades}. These analyses are strongly motivated by our lack of knowledge about the origin and propagation of neutrinos, and by the fact that they can even be a powerful tool to searching for  beyond the standard model physics \cite{markus}. The increasing in data collection time at the IceCube and the expectation of a next generation of neutrino observatories in the coming decades \cite{icecube-gen2},  makes possible to search for subleading effects and rare processes not yet observed. In this context, several authors have pointed out the current theoretical uncertainties present in the ingredients necessary to estimate the observed events \cite{soni,palomares1,palomares2,Klein:2019nbu,Goncalves:2021gcu,Goncalves:2022uwy}.

\begin{figure}[!t]
	\centering
	\begin{tabular}{ccccc}
	   \includegraphics[width=0.25\textwidth]{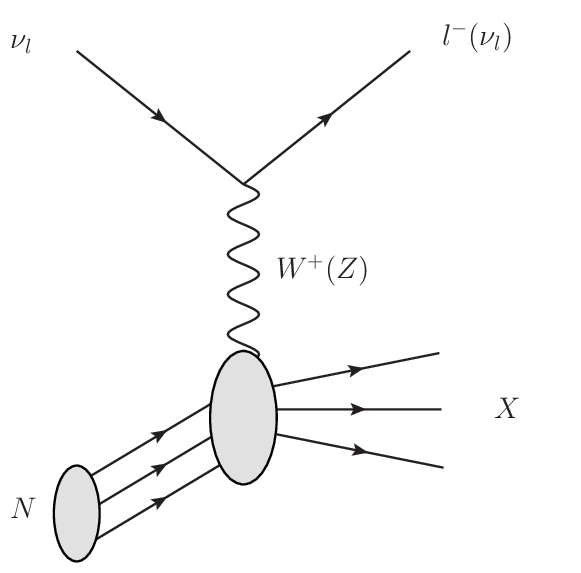} & 
          \includegraphics[width=0.25\textwidth]{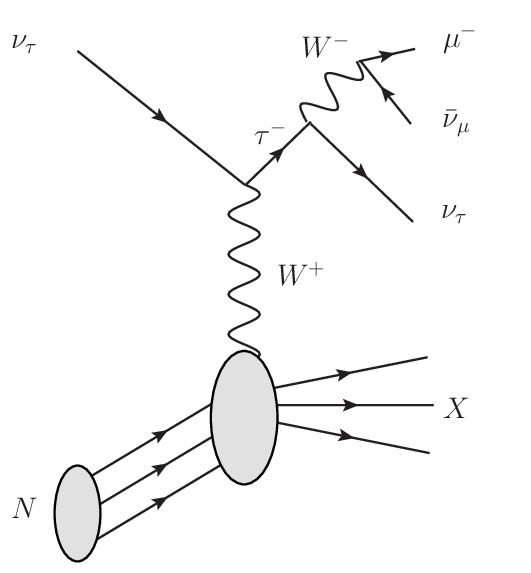} & 
          \includegraphics[width=0.25\textwidth]{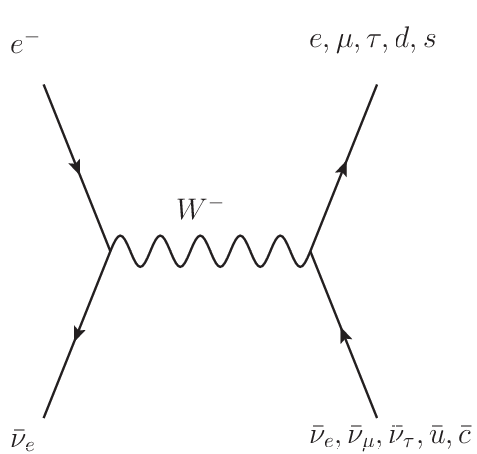} \\
          (a) & (b) & (c) \\
          \includegraphics[width=0.3\textwidth]{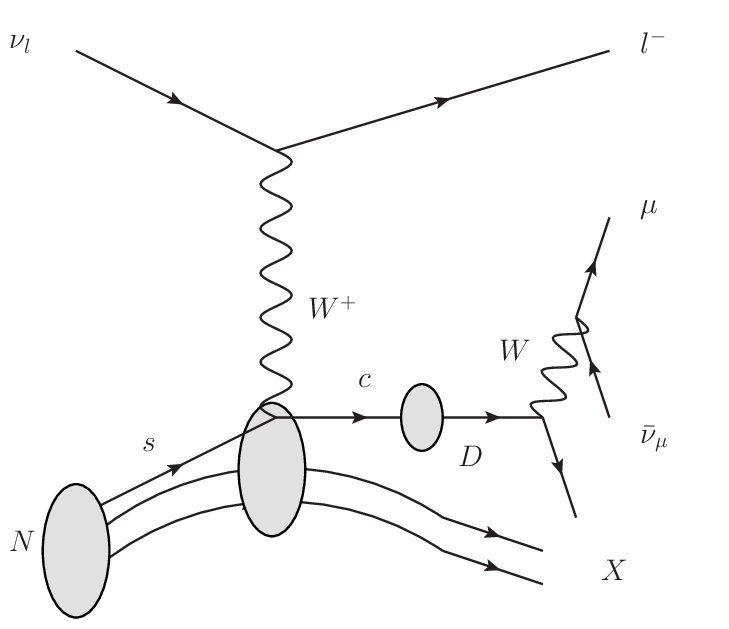} &
          \includegraphics[width=0.3\textwidth]{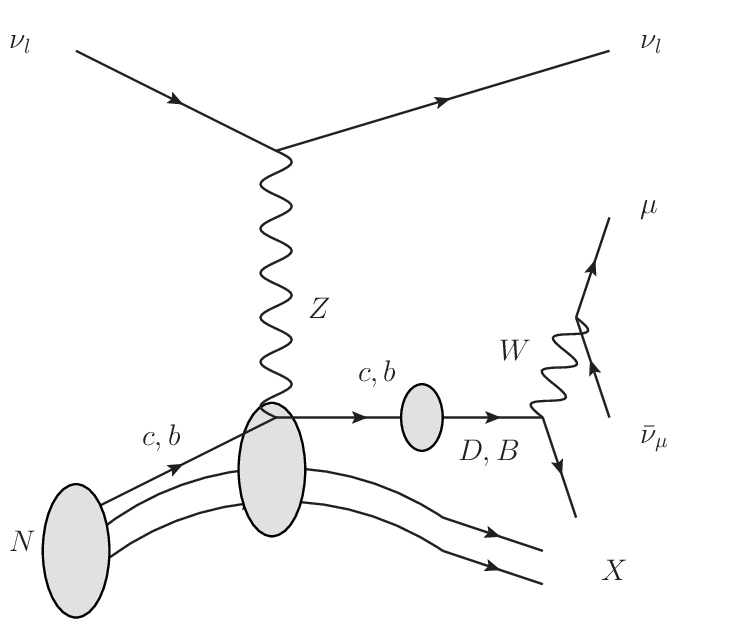} & 
          \includegraphics[width=0.3\textwidth]{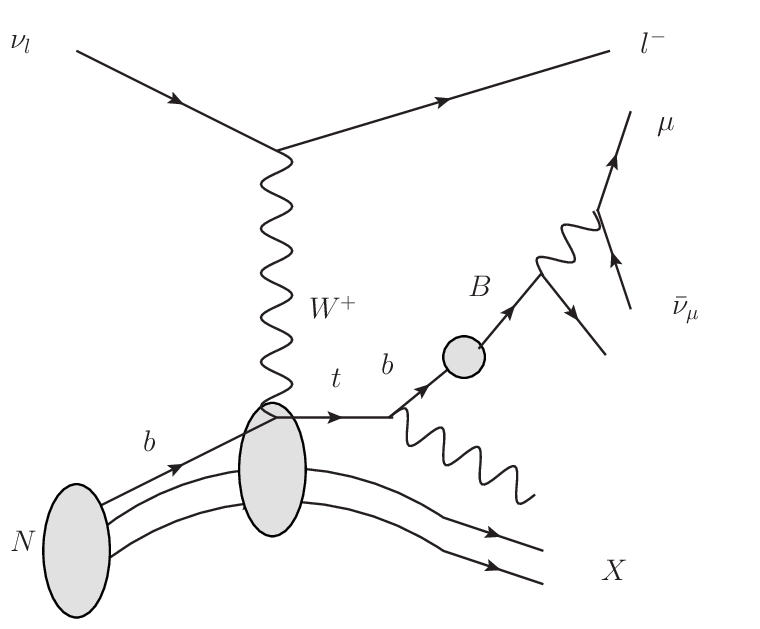} \\
          (d) & (e) & (f) \\
          \includegraphics[width=0.3\textwidth]{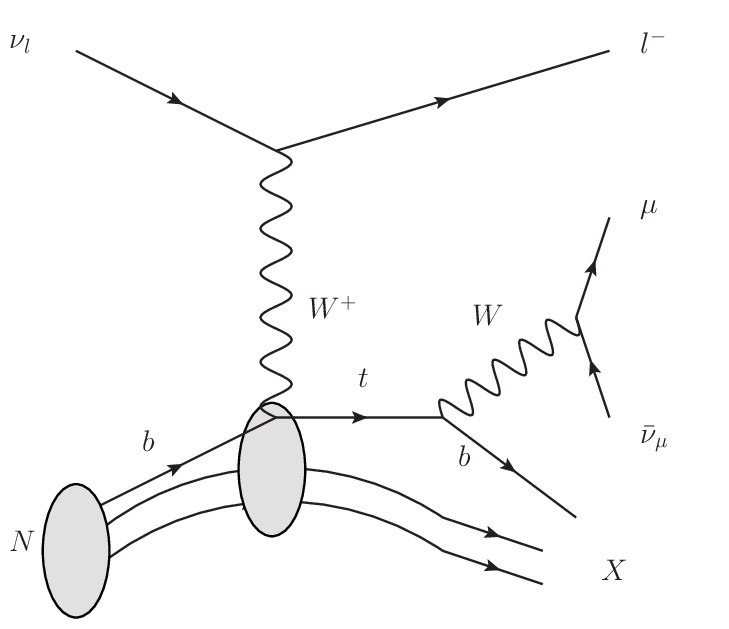} &
          \includegraphics[width=0.3\textwidth]{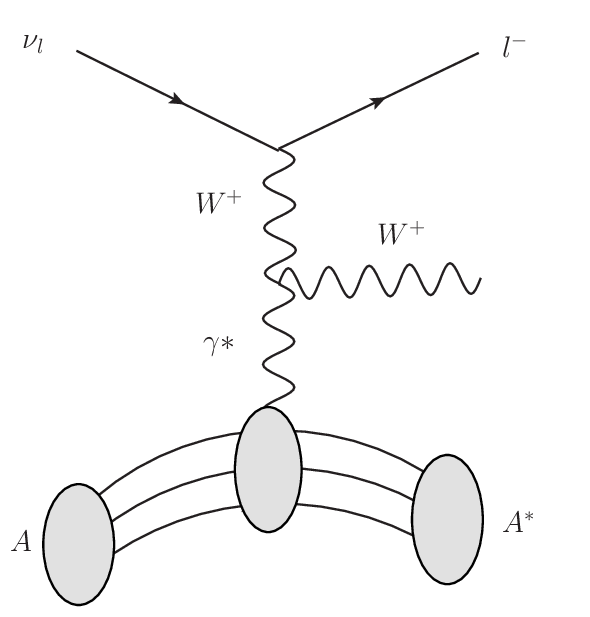} & 
          \includegraphics[width=0.3\textwidth]{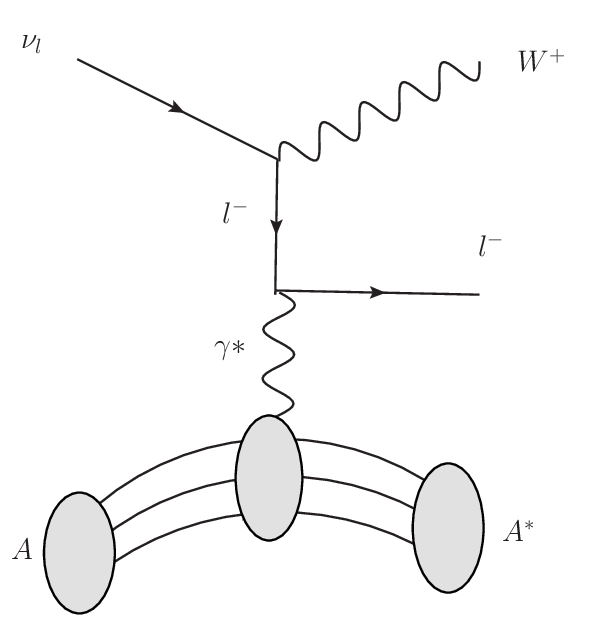} \\
          (g) & (h) & (i)
	\end{tabular}
\caption{ Typical diagrams for the interaction of neutrinos in different channels that give rise to muons in the final state, which can be identified as tracks at the IceCube observatory. }
\label{fig:diagramas}
\end{figure}

Neutrino detection is indirect in IceCube, being observed by Cherenkov light emitted by relativistic charged particles produced in each process  \cite{HESE2}. In particular, for High Energy Start Events (HESE),  there are three distinguishable topologies: cascade, double cascade and starting track (for simplicity we will just call it tracks here). Cascades are approximately spherical localized energy deposits. They arise from electronic neutrino interactions via charged current (CC) and all neutral current (NC) interactions. In addition,  CC interactions of tauonic neutrinos also contribute to cascade events when the tau decay into hadrons or in an electron occurs sufficiently near to the first interaction vertex. 
Tracks are events in which there is a very energetic muon in its final state, which generally travels several kilometers before decaying, leaving the instrumented volume of the observatory. Tracks are about $25\%$ of the events observed in HESE. Its production occurs mainly in a CC interaction of a muon (anti)neutrino [Fig. \ref{fig:diagramas} (a)] and from the tau decay produced in tau (anti)neutrino CC interactions [Fig. \ref{fig:diagramas} (b)]. In addition, the Glashow resonance process [Fig. \ref{fig:diagramas} (c)], also contributes to the production of an energetic muon. The contribution of these channels is usually taken into account for calculating the number of track events at the IceCube \cite{soni,palomares1,palomares2,Goncalves:2021gcu,Goncalves:2022uwy}.   Muons in the final state can also be generated in the decay of heavy mesons produced in charged and neutral interactions induced by neutrinos of all flavors 
[Figs. \ref{fig:diagramas} (d) - (f)], in the top decay [Fig. \ref{fig:diagramas} (g)] and in the $W$ - production process [Figs. \ref{fig:diagramas} (h) and (i)]. In general, the associated contribution for track events is assumed to be subleading and is disregarded in the phenomenological studies. One of the goals of this paper, is to estimate the contribution of these subleading channels for the number of track events at the IceCube, as well as to investigate its impact on the determination of the astrophysical neutrino flux.
Another goal, is to estimate the ratio between the energy associated with the cascade and its sum with the total reconstructed energy of the muon, which defines the reconstructed average inelasticity for the subleading processes.
In principle, subleading channels are expected to divide the neutrino energy between cascade events and the final muon differently than in the case of CC $\nu_\mu$ interactions and, consequently, the  inelasticity could be used to discriminate the distinct contributions for track events. Our analysis is motivated by the  study performed by the IceCube Observatory \cite{IceCube:2018pgc}, which have demonstrated that tracks can be used to investigate the inelasticity of the observed events.



Our paper is organized as follows: in the next section,  we present a brief review of the formalism needed to calculate the contribution associated with the diagrams presented in Fig. \ref{fig:diagramas} for the number of events at the IceCube. Furthermore,  the reconstructed average inelasticity associated with track events is discussed. In section \ref{sec:res} we present our results for the energy dependence of the cross - sections associated with the leading and subleading channels, for the reconstructed mean inelasticity and for the number of events at the IceCube. We also show the impact of additional channels on the description of the astrophysical neutrino flux, considering the current HESE data for tracks. Finally, in section \ref{sec:sum}, we summarize our main results and conclusions.

\section{Formalism}

The main observable in Neutrino Observatories is the number of neutrino-induced events, which is given by 
\begin{eqnarray}
\mathrm{d}N_{events} = 
T \sum_{\alpha, l}
N_{effe,\alpha}(E_{\nu})\times 
\Phi_{\nu_l} (E_{\nu})\times
\sigma _{\nu _l \alpha}(E_\nu) \times
T_{\nu_l}(E_\nu , \theta_{z})\,
{\mathrm{d}(E_{vis})\, \mathrm{d}\Omega}\, ,
\label{eq:eventos}
\end{eqnarray}
where $T$ is the observatory exposure time for data collection, $N_{effe, \alpha}$ is the effective number of targets of type $\alpha$ for scattering, $\Phi_{\nu_l}$ is the astrophysical neutrino flux, $\sigma_{\nu_l \alpha} (E_\nu)$ is the neutrino-target cross - section, and $T_{\nu_l}(E_\nu, \theta_z)$ is the coefficient of transmission of  neutrinos of flavor $l = e, \mu, \tau$ when crossing the Earth with zenith angle $\theta_z$. In our analysis, the effective number of targets in the observatory has been obtained from \cite{palomares2}, where this number is presented as a function of the energy deposited in the interaction process. Moreover, $T_{\nu_l}(E_\nu, \theta_z)$ is calculated following Ref. \cite{francener1}, assuming the PREM profile density for the Earth and taking into account the flux regeneration associated with  neutral current interactions and tau lepton decay.


An important quantity present in Eq. (\ref{eq:eventos}) that we need to describe is the astrophysical neutrino flux, $\Phi_{\nu_l} (E_\nu)$. The origin of this flux is not yet well understood, but the data points to an extragalactic origin, even though IceCube has identified only two possible point sources \cite{blazar1,blazar2,sn1068}. In our analysis, we assume that this flux is equally distributed between neutrinos and antineutrinos, as well as between the three flavors. We also assume that this flux is isotropic and parameterized by a power law,
\begin{eqnarray}
\Phi_{\nu_l} (E_\nu) = \Phi_0 \left( \dfrac{E_\nu}{ 100\;\mathrm{TeV}}\right) ^{-\gamma} \; 10^{-18} \mathrm{GeV}^{-1} \mathrm{s}^{-1} \mathrm{sr}^{-1}\mathrm{cm}^{-2}\, .
\label{eq:fluxo}
\end{eqnarray}
The flux normalization parameters, $\Phi_ 0$, and spectral index, $\gamma$, will be obtained using a maximum likelihood to  our prediction fit compared to the current IceCube data.

At high neutrino energies, the  neutrino - hadron interaction is dominated by the  deep inelastic scattering (DIS) processes, represented in Fig. \ref{fig:diagramas} (a). This interaction is characterized by the exchange of a virtual boson ($W^{\pm}$ or $Z$) with a four-momentum $q^{2} = -Q^{2} < -1\, \mathrm{GeV}^{2} $ and by the fragmentation of the nucleon target, generating a hadronic final state with an invariant mass $W > 1.4\, \mathrm{GeV}$. In a CC interaction, characterized by the exchange of a $W^{\pm}$, the (anti)neutrino $\nu_{l}$ with energy $E_{\nu}$ becomes its corresponding (anti)charged lepton, $l = e, \mu , \tau$, with energy $E_{l}$. In contrast, in a NC interaction, where a $Z$ boson is exchanged, the final state is characterized by a neutrino of the same flavor as the incident one. Usually the DIS is described in terms of the gauge boson virtuality $Q^2$, the Bjorken-$x$ variable (defined by $x = Q^2/2p\cdot q$ with $p$ being the four-momentum of the target nucleon), and the inelasticity $y = p\cdot q / p\cdot k$ (with $k$ being the initial neutrino four-momentum). In the nucleon's rest frame, $y$ becomes $(E_\nu - E_l)/E_\nu$, and is then interpreted as the fraction of the neutrino's initial energy transferred to the hadronic state. In the context of the variables described above, the double differential CC neutrino-nucleon cross - section is given by \cite{paschos,reno}
\begin{eqnarray}
\begin{aligned}
\frac{\mathrm{d}\sigma^{\nu_{l}N}}{\mathrm{d}x\mathrm{d}y}=
\frac{G_F^{2}m_N E_\nu}{\pi}
\left(
\frac{M_W^2}{Q^2+M_W^2}
\right)^2
\left\{
\left(
y^2x+\frac{m_l^2y}{2E_\nu m_N}
\right)F^{CC}_1(x,Q^2)+\right. 
\left(
1-y-\frac{m_l^2 y}{4E_\nu^2}-\frac{m_Nxy}{2E_\nu}
\right)F^{CC}_2(x,Q^2)+ \\
 + \left(
xy-\frac{xy^2}{2}-\frac{m_l^2y}{4E_\nu m_N}
\right)F^{CC}_3(x,Q^2)+ 
\left. \frac{m_l^{2} (m_l^{2}+Q^2)}{E_\nu^2 m_N^2 x}F^{CC}_4(x,Q^2)
-\frac{m_l^2}{E_\nu m_N}F^{CC}_5(x,Q^2)
\right\} \, ,
\end{aligned}
\label{eq:cs}
\end{eqnarray}
with $G_{F}$ being the Fermi's constant, $m_l$ the charged lepton mass,  $M_{W}$ the  $W$ - boson mass and $F^{CC}_{i}$ are the CC nucleon structure functions, which can be expressed in terms of the parton distribution functions (PDFs) of the nucleon. For the PDFs, we will assume the CT14 parametrization \cite{ct14}.  The corresponding expression for the case of an incoming antineutrino can be obtained by reversing the sign of the $F_3$ term. A similar equation can be derived for NC scattering, differing by the mass of the lepton in the final state and by the structure functions \cite{paschos}. The total cross - section can be obtained from Eq. (\ref{eq:cs}), integrating into $x$ and $y$ in the kinematically allowed ranges \cite{paschos}.
In order to include the contribution associated with the top production in CC neutrino - bottom interactions, we  consider the approach described in detail in Ref.  
\cite{barger1}.


In what follows, we will focus on track events, where an energetic muon is present in the final state. Such a muon can be directly generated in the leptonic vertex of DIS through a  CC $\nu_{\mu}$ - hadron interaction [Fig. \ref{fig:diagramas} (a)], with its contribution determined by Eq. (\ref{eq:cs}). The  muon can also be created in the decay of a tau lepton generated in the leptonic vertex of DIS through a  CC $\nu_{\tau}$ - hadron interaction [Fig. \ref{fig:diagramas} (b)]. If the decay occurs near of the first leptonic vertex, the event is not classified as a double bang event.  Such a contribution can also be estimated using Eq. (\ref{eq:cs}) and taking into account of the tau decay. In addition, a muon can also be created as a subproduct of the particles produced in the hadronic vertex of DIS for CC and NC neutrino - hadron interactions. In particular, an energetic muon is expected to be produced in the decay of heavy mesons as, e.g.,  $D$ and $B$ mesons [Figs. \ref{fig:diagramas} (d) - (f)], or in the decay of a top quark [Fig. \ref{fig:diagramas} (g)]. In our analysis, we have estimated the associated contributions, calculating the DIS cross - sections for the heavy meson and top production, and taking into account of its decays into a muon. As heavy mesons have lifetimes of the order of $10^{-12}\,s$,  we consider that they decay before interacting. The hadronization and decays will be simulated using the \texttt{Pythia6} Monte Carlo generator \cite{Sjostrand:2006za}.


In addition to the DIS channels, muons in the final state can be generated through the  Glashow resonance and $W$ - boson production processes.  In particular, in the case of the Glashow resonance process, the annihilation of an electronic antineutrino with an electron into a $W$ - boson [Fig. \ref{fig:diagramas}(b)], can  directly generate muons or indirectly from the decay of taus and charm quarks. In our analysis, we include both possibilities. On the other hand,  in the $W$ - production process [Figs. \ref{fig:diagramas}(h) and (i)], one has a $\nu_l\,\gamma^* \rightarrow W^{\pm} + l^{\mp}$ interaction, where the virtual photon is emitted by the target, which can remain intact or dissociate. Such a channel is of particular interest for track events, given that muons are produced directly in the case of interaction induced by $\nu_\mu$, and also come from the decays of tau and  $W$ - boson produced in the interaction. Such contributions are estimated following Refs. \cite{beacom1,beacom2}.


Finally, the last quantity that we describe here, necessary for the calculations of our interest, is the average inelasticity, $\langle y\rangle$.
The average inelasticity in the neutrino-nucleon DIS interaction is  defined by
\begin{eqnarray}
\begin{aligned}
\langle y (E_\nu) \rangle = 
\frac{\int \mathrm{d}y \int \mathrm{d}x \, y \,\frac{\partial\sigma^{\nu_{l}N}}{\partial x\partial y}} 
{\int \mathrm{d}y \int \mathrm{d}x \frac{\partial\sigma^{\nu_{l}N}}{\partial x\partial y}} \,\,.
\end{aligned}
\label{eq:inelasticity}
\end{eqnarray}
This quantity determines the energies carried by the hadronic state ($E_h = y E_{\nu}$) and by the lepton arising from the vertex with the neutrino ($E_l = (1-y) E_{\nu}$). It is important to emphasize that the detector efficiencies in the reconstruction of $E_h$ and $E_{\nu}$ are not identical. In our analysis, the equivalent electromagnetic  energy of a hadronic cascade of energy $E_{h}$ will be assumed to be equal to $E_{h}f_{h}(E_h)$, where $f_{h} (E_h)$ is described by the parametrization proposed in Ref. \cite{Gabriel:1993ai,Kowalski} for the IceCube Observatory. On the other hand, as highly energetic muons deposit only a fraction of their energy in the detector, we will use the parameterization proposed in Ref. \cite{palomares2} to quantify this fraction in the calculation of the energy deposited in each event, necessary in the case of HESE.

One has that Eq. (\ref{eq:inelasticity}) directly reports the energy of the final muon and hadronic cascade only for muons arising from CC $\nu_\mu$ - hadron interactions. For the subleading processes, one has to take into account that the muons arise from the decay of quarks, heavy mesons or $W$ - bosons. For these processes, we can define the 
reconstructed average inelasticity as the ratio between the energy seen as a cascade in the event, $E_{casc.}$, and the sum of $E_{casc.}$ with the total reconstructed energy of the final muon, which is identical to $\langle y (E_\nu) \rangle$ for $\nu_\mu$ - hadron interactions.
For the process represented in Fig. \ref{fig:diagramas}(b),
we will have that $E_{casc.} = y E_\nu f_h$, while the energy of the final muon will be $E_\mu = E_\nu (1- y)z_{\tau \rightarrow \mu}$, where $z_{\tau \rightarrow \mu}$ is the fraction of the tau energy that remains with the final muon. On the other hand, for the production of a heavy meson induced by  CC $\nu_e$ - hadron interaction, which is a particular case of the diagram represented in Fig. \ref{fig:diagramas} (d), 
$E_{casc.}$ is the sum of the electron energy, $E_\nu (1-y)$, with the energy transferred to the target nucleon subtracted from the lepton energies, $yE_\nu f_h (1-z_{h \rightarrow \mu} - z_{h \rightarrow \bar{\nu}_\mu})$. Moreover, the muon energy  is $y E_\nu f_h z_{h \rightarrow \mu}$, where  $z_{h \rightarrow \mu}$ and $z_{h \rightarrow \bar{\nu}_\mu}$ are the energy fractions of the hadronic part of the final state that remains with the muon and the neutrino, respectively, after of semileptonic decay. Similar analyzes can be performed for the other processes in order to estimate the associated values of $E_{casc.}$ and $E_\mu$.


\section{Results}
\label{sec:res}

\begin{figure}[t]
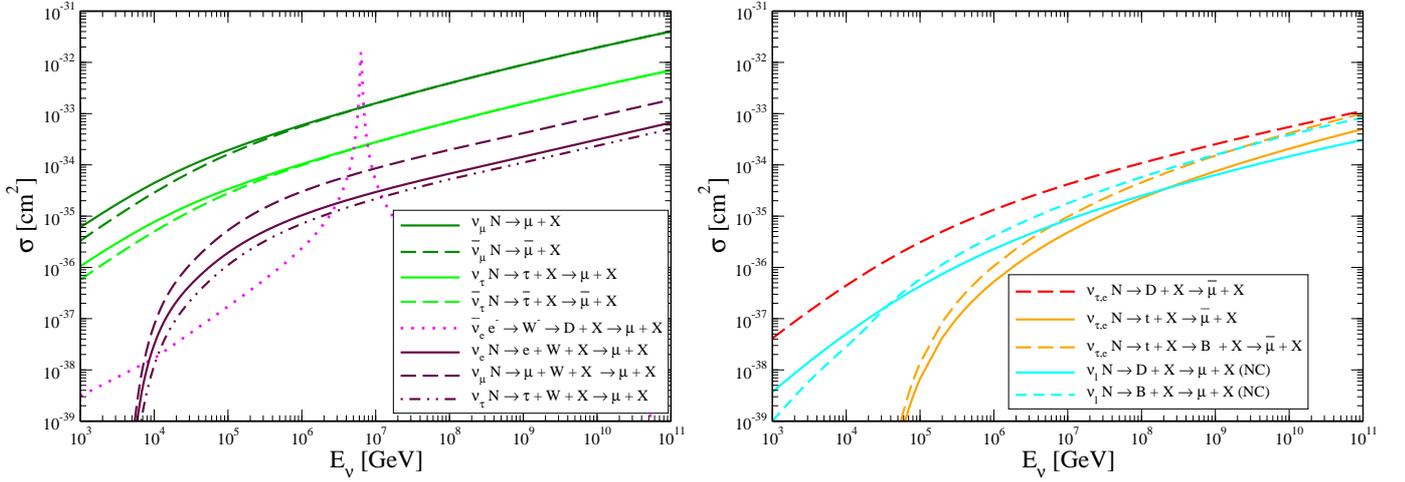

	\centering
	\begin{tabular}{ccc}
	   \includegraphics[width=0.5\textwidth]{sigma_MuonFinalState_1.eps} &
	   \includegraphics[width=0.5\textwidth]{sigma_MuonFinalState_2.eps}
	\end{tabular}
\caption{ Energy dependence of the neutrino and antineutrino cross - sections per nucleon target (per electron, for the Glashow resonance) for the production of an energetic muon in the final state. On the left panel we present the cross - sections for muons arising from lepton vertices and from the decay of the $W$ - boson, while on the right panel the muons arise from the decay of heavy mesons and  top quark. \vspace{1cm}}
\label{fig:sigma}
\end{figure}

Initially, we will investigate the energy dependence of the cross - sections  per nucleon (per electron, for Glashow resonance)  for the processes that generate energetic muons in the final state and that can be identified as tracks. The results will be estimated requiring that the energy of the muon in the final state is greater than the IceCube threshold of 70 GeV. Our predictions are presented in Fig. \ref{fig:sigma}. On the left panel, we present the cross - sections for muons arising from lepton vertices and from the decay of the $W$ - boson, while on the right panel we present the predictions associated with the processes where the muons arise from the decay of heavy mesons and top quarks. 
One has that the cross - section is dominated by processes induced by muon neutrinos, except at the peak of the Glashow resonance, $E_\nu \approx$ 6.3 PeV. In this figure, we show that several channels of electronic and tauonic neutrino interactions make non-negligible contributions to track signals.  In particular, the contributions associated with the decay of taus and heavy mesons become important with the increasing of the neutrino energy. Moreover, in agreement with the results derived in Refs. \cite{beacom1,beacom2},  the cross - section for the $W$ - boson production induced by muon neutrinos also contributes significantly for the muon production in the kinematical range probed by the IceCube. On the other hand, if the process is induced by electronic and tauonic neutrinos, its contribution is smaller by almost one order of magnitude, which is associated with the fact that in this case the muons only arise from the decay of their products.
We also have that the CC cross - sections for  the
$\nu_{\mu ,\tau}N$ and $\bar{\nu}_{\mu ,\tau}N$ processes are similar at high energies, but different for energies smaller than $10 ^{6}$ GeV due to the term $F_3$ in Eq. (\ref{eq:cs}), which is sensitive to the valence quark content of the hadrons. In contrast, for the production of heavy mesons and top quarks one has that   the cross - sections for processes induced by neutrinos and antineutrinos are equal, since they are dependent on the sea quark content. Similarly, for the W - boson production processes, we also have equal cross - sections induced by neutrinos and antineutrinos of a given flavor.

\begin{figure}[t]
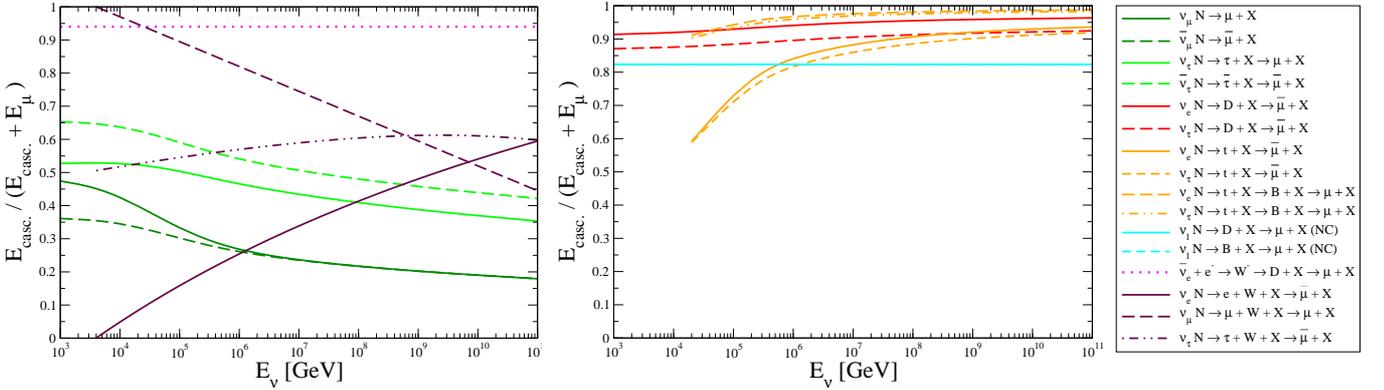

	\centering
	\begin{tabular}{ccc}
	\includegraphics[width=0.4\textwidth]{Y_med_1.eps}
	\includegraphics[width=0.59\textwidth]{Y_med_2.eps}
			\end{tabular}
\caption{ Reconstructed average inelasticity as a function of neutrino energy for different channels of muon production. }
\label{fig:Ymedio}
\end{figure}

In order to determine the contribution of the subleading processes for the track events, in addition to the energy dependence of the associated  cross - sections, it is important to know how each channel deposits equivalent electromagnetic energy in the detector, and how this energy is divided between the visible energy deposited in the cascade and  the energy deposited by the muon. In what follows, we will present our predictions for the reconstructed average inelasticity, given by the ratio $E_{casc.} / (E_{casc.}+E_{\mu})$, where $E_{casc.}$ is the electromagnetic equivalent  deposited energy by all particles except for the muon and $E_{\mu}$ is the reconstructed muon energy. Such quantities are calculated for the different processes as explained in the previous section. The results are shown in 
Fig. \ref{fig:Ymedio}. 
One has that  muons arising from heavy meson and top  decays (right panel) are on average much less energetic than $E_{casc.}$, generally having less than 20$\%$ of its energy. Conversely, for interactions initiated by muon neutrinos (left panel), the ratio decreases from $0.47$ to $E_{\nu}=10^{3}\, \mathrm{GeV}$ for less than $0.2$ in $E_{\nu}=10^{11}\, \mathrm{GeV}$, which means that a very energetic muon is produced at high energies in this process.  For muons generated in tau decays, one has that the ratio is approximately $0.65$ ($0.53$) in $E_{\nu} = 10^{3}\, \mathrm{GeV}$ and drops up to $0.42$ ($0.35$) at $E_{\nu} = 10^{11}\, \mathrm{GeV}$ for processes induced by antineutrinos (neutrinos). Therefore, at high energies, the muons generated in the tau decays are $\approx 15\%$ less energetic than those directly produced in $\nu_\mu N$ interactions.

For the $W$ - boson production processes (left panel), the behavior of the reconstructed average inelasticity  is strongly dependent on the flavor of the incident neutrino. One has that for  $\nu_e$ interactions, the ratio starts from  zero at the $W$ - boson production threshold and grows to 0.6 at $E_\nu \approx 10^{11}\, \mathrm{GeV}$. On the other hand, an opposite behavior is observed for events started by $\nu_\mu$, where it starts from 1 and decreases to 0.4 at the highest energy considered. These opposite behaviors for interactions initiated by $\nu_e$ and $\nu_\mu$ are explained by the origin of the muons in the final state: in the muon neutrino interaction, the muon comes from the vertex of the initial neutrino, while in the electronic neutrino interaction the muon comes from the decay of the $W$ - boson produced, which on average is much more energetic than the lepton produced in the IceCube energy regime. Our results also indicate that the reconstructed average inelasticity for the $W$ - boson production induced by tau neutrinos has an intermediate behavior between those predicted for the  two other neutrino flavors, which is associated with the fact  that similar amounts of muons come from tau and $W$ - boson decays. As a summary, Fig. \ref{fig:Ymedio} shows that the subdominant channels imply an  energy distribution between muons and cascades very different from that for charged current $\nu_\mu N$ interactions. As a consequence, if these subleading processes cannot be separated, its contributions have a direct  impact on the determination of the average inelasticity.

\begin{figure}[t]
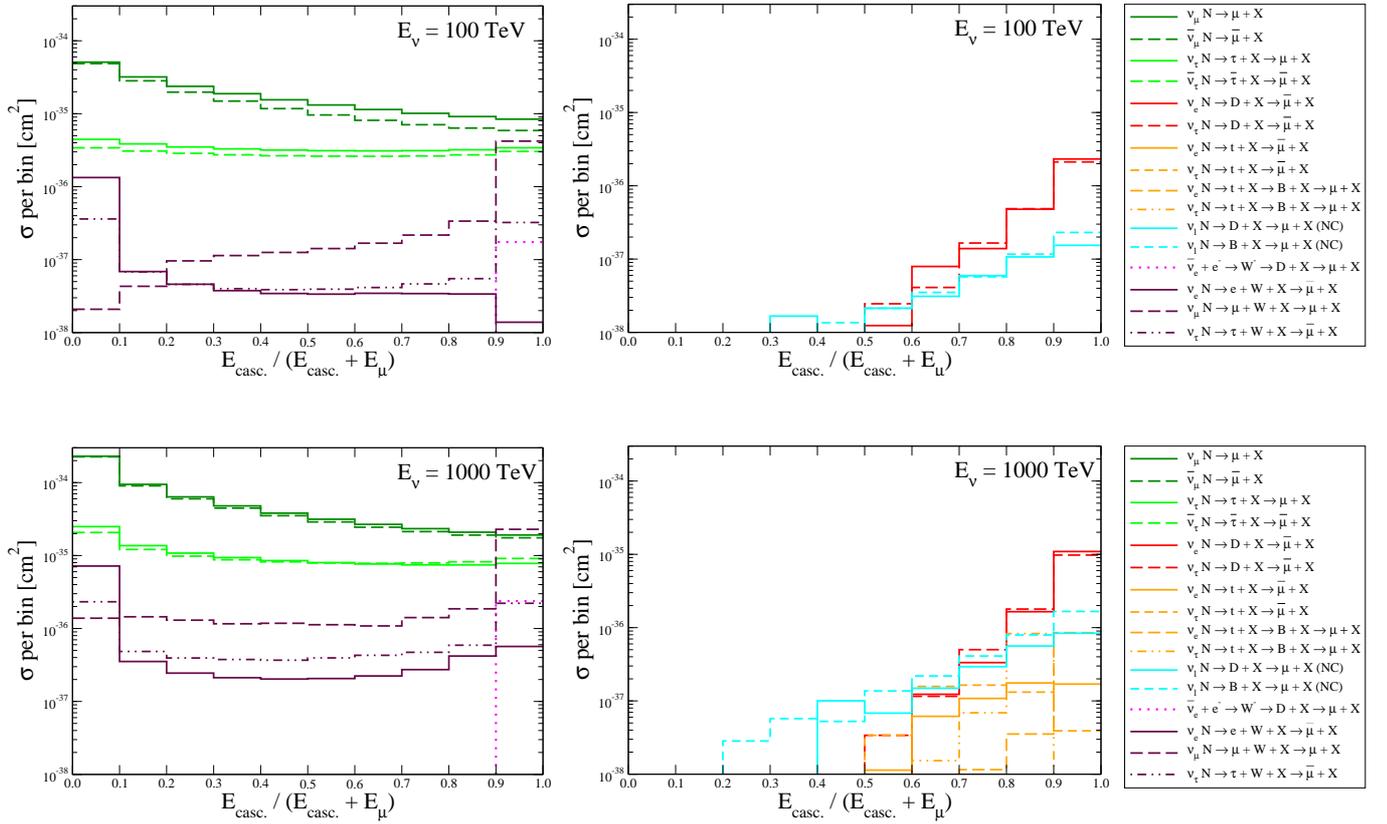

	\centering
	\begin{tabular}{ccc}
    \includegraphics[width=0.4\textwidth]{histograma_100TeV_1.eps} &
    \includegraphics[width=0.59\textwidth]{histograma_100TeV_2.eps} \\
    \,\,\, & \,\,\, \\
     \,\,\, & \,\,\, \\
    \includegraphics[width=0.4\textwidth]{histograma_1000TeV_1.eps} &
    \includegraphics[width=0.59\textwidth]{histograma_1000TeV_2.eps}
			\end{tabular}
\caption{ Cross - section as a function of the reconstructed average inelasticity for several channels with muons in the final state. We consider neutrino energy of 100 TeV (upper panels) and 1000 TeV (lower panels). }
\label{fig:histograma}
\end{figure}

For completeness of our analyses, in Fig. \ref{fig:histograma} we present the predictions for  the cross - sections as a function of the reconstructed average inelasticity for the same channels described in the previous figures, assuming that the  incident neutrino has an energy of $10^{5}\, \mathrm{GeV}$ (upper panels) or $10^{6}\, \mathrm{GeV}$ (lower panels). As expected from Fig. \ref{fig:Ymedio}, the contributions associated with the heavy meson and top decays (right panels) imply muons much less energetic than the equivalent electromagnetic energy observed coming from the other scattering products. In $W$ - boson production processes (left panels), we have a maximum in the first and last bin for interactions started by $\nu_e$ and $\nu_\mu$, respectively. In the case of interactions induced by tauonic neutrinos, we have two local maxima, one related to the decay of tau, and the other to the decay of the $W$ - boson. Conversely, the cross - section for muon neutrinos has its maximum in the first bin and its minimum in the last bin, indicating more energy with the muon coming from the leptonic vertex than from the hadronic products. The cross - section is larger in all bins for muon neutrinos compared to all channels when we compare them individually. However, especially at higher energies of the incident neutrino,  the summed subdominant channels in the last bin make a more significant contribution.

\begin{figure}[t]
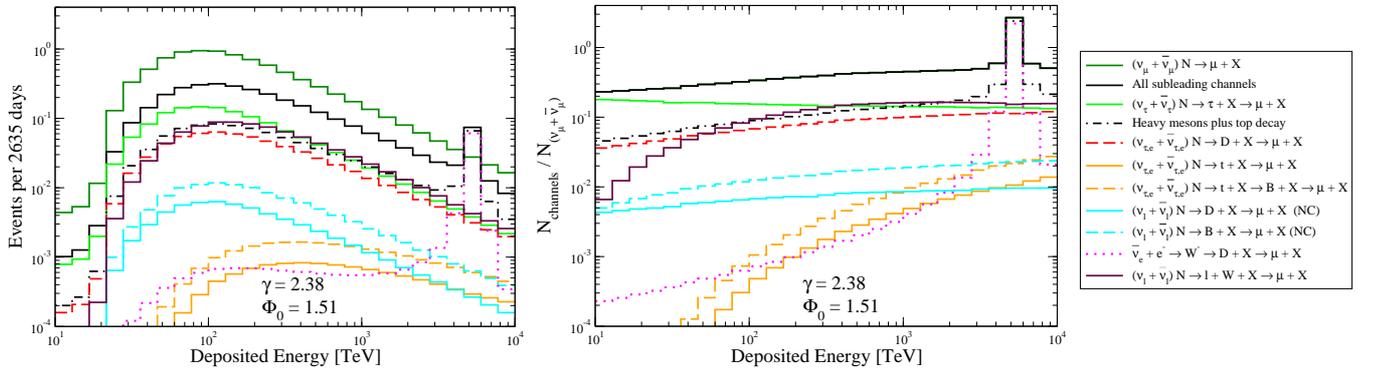

	\centering
	\begin{tabular}{ccc}
	\includegraphics[width=0.38\textwidth]{N_events_tracks_chanels.eps} &     \includegraphics[width=0.6\textwidth]{ratio_contributions.eps}
			\end{tabular}
\caption{ {\bf Left panel:} Tracks events at IceCube for different channels as a function of deposited energy. We consider 7.5 years of exposure and an astrophysical flux parametrized with $\gamma = 2.38$ and $\Phi_0 = 1.51$. {\bf Right panel:} Ratio between the contributions to tracks associated to the subleading channels and that for $\nu_\mu$ CC interactions. }
\label{fig:events_channels}
\end{figure}

In order to  quantify the importance of tracks coming from subdominant channels, in Fig. \ref{fig:events_channels} (left panel) we present of the number of tracks events, expected in the IceCube Observatory considering a period of 7.5 years,  as a function of the electromagnetic equivalent deposited energy. In this calculation, we have assumed  $\gamma = 2.38$ and $\Phi_0 = 1.51$ in the astrophysical neutrino flux, which is the best fit for the last analysis of tracks in IceCube observatory \cite{tracksTotal}. The number of events is dominated by muon neutrino interactions, except at the peak of the Glashow resonance. In Fig. \ref{fig:events_channels} (right panel) we show the ratio between the contributions of the subdominant channels and the contribution of muon neutrinos. At the lower deposited energies ($10^{4}-10^{5}\, \mathrm{GeV}$), the main subdominant contribution is from taus decay, but with increasing deposited energy this channel is overcome by contributions from $W$ - boson production and heavy meson and top decays, with the heavy meson decay contribution being the most important at higher deposited energies. The total subdominant contributions (sum of events arising from tau, W - boson and heavy meson and top decays) represents more than $20\%$ of the observed tracks astrophysical neutrino events in all bins present in Fig. \ref{fig:events_channels}, and in the bin with maximum contribution from the Glashow resonance it reaches $200\%$ of the contribution of CC muon neutrino interactions.

\begin{figure}[t]
	\centering
	\begin{tabular}{ccc}
	\includegraphics[width=0.6\textwidth]{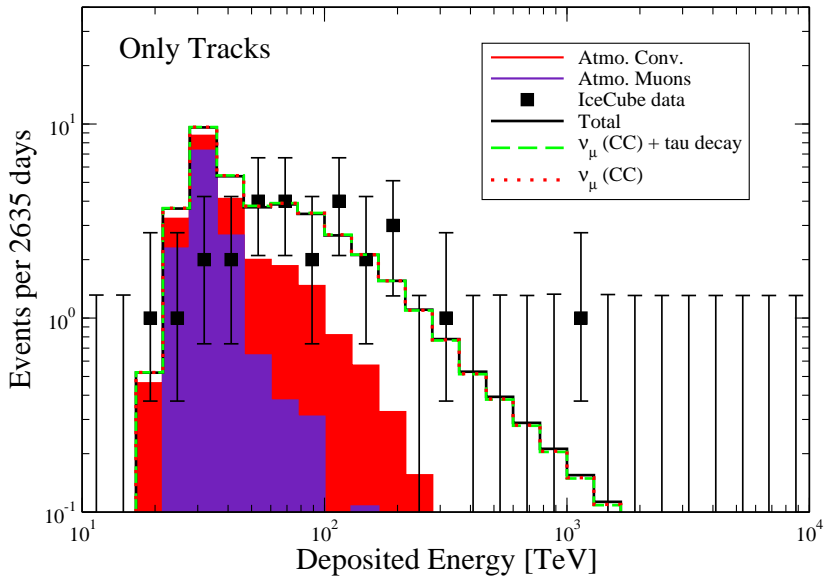}
	\end{tabular}
\caption{ Tracks events at IceCube considering 7.5 years. Data from  HESE for tracks. }
\label{fig:events}
\end{figure}

Finally, in Fig. \ref{fig:events} we present our best fit to the IceCube track events data considering a period of 2635 days, for events initiated in the instrumented volume of the observatory. We use the atmospheric neutrino and muon backgrounds given in \cite{Schneider:2020tyd}. We show our best fit considering three different configurations: total events added over all channels described previously; considering the contributions of events initiated by CC muon neutrino interactions and by the decay of taus into muons; and considering only the contributions of CC muon neutrino interactions. The likelihood analysis was carried out to obtain the best fit in each configuration by adjusting the flux parameters ($\gamma , \Phi_0$) and also normalizing the backgrounds involved and adjusting the energies observed using the Pull method. In our analysis, we used bins 7-19, as it is a region with data where astrophysical neutrinos dominate over the  atmospheric backgrounds. Given the low statistics of the current data, the Pull method does not modify the best background normalization parameters, nor does it shift the visible energy from the central value. Conversely, the parameters of the astrophysical flux are modified when we consider the different contributions for the analyzed events. Considering all the channels studied, the contributions from muon neutrinos plus tau decay, and considering only muon neutrinos, we obtain for $\gamma$ and $\Phi_0$ the following results: (2.57, 2.16), (2.55, 2.52) and (2.55, 2.90), respectively. The spectral index is very weakly modified when we consider more track production channels. This is a consequence of the energy dependence of the contributions being similar in the analyzed region. However, the flux normalization changes significantly: it decreases by $13\%$ and $25\%$ when we insert the contributions from tau decay and $W$ - boson production plus heavy meson and top decays, respectively.


\section{Summary}
\label{sec:sum}

Track signals account for more than $25\%$ of the total number of events observed at HESE. The dominant amount of these events comes from charged current interactions of muon neutrinos. In this work, we have shown that signals arising from charged current interactions of tauonic neutrinos, $W$ - boson production and decays of heavy mesons and top quarks make up a non - negligible contribution for track events, which can be greater than $20\%$ of the events induced by $\nu_\mu$ at the DIS regime.
We also have shown that these subdominant channels contribute to events with greater reconstructed average inelasticity than in the case of traces arising from CC interactions induced by $\nu_\mu$. This makes it feasible to search for the contributions of these signals in IceCube and similar observatories.
Furthermore, our results indicate that the inclusion of track signals from non-neutrino muonic CC events is necessary to describe the parameters of the astrophysical flux, especially those referring to the parameter of its normalization.


\begin{acknowledgments}

R. F. acknowledges support from the Conselho Nacional de Desenvolvimento Científico e Tecnológico (CNPq, Brazil), Grant No. 161770/2022-3. V.P.G. was partially supported by CNPq, FAPERGS and INCT-FNA (Process No. 464898/2014-5). D.R.G. was partially supported by CNPq and MCTI.

\end{acknowledgments}

\hspace{1.0cm}

\end{document}